\documentclass{jfm}
\usepackage{graphicx}
\usepackage{amsmath}
\usepackage{amssymb,mathrsfs}

\usepackage[normalem]{ulem}
\usepackage{color}
\usepackage{soul}

\begin{document}

\shorttitle{Holes and cracks in rigid foam films} 
\shortauthor{P. C. Petit et al.} 

\title{Holes and cracks in rigid foam films}

\author
 {
P. C. Petit \aff{1},
  M. Le Merrer \aff{1}
  \and 
 A.-L. Biance \aff{1}
  \corresp{\email{anne-laure.biance@univ-lyon1.fr}},
  }

\affiliation
{
\aff{1}
Institut Lumi\`ere Mati\`ere, Universit\'e de Lyon, UMR5306 Universit\'e Lyon 1-CNRS, 69622 Villeurbanne, France
}

\maketitle

\begin{abstract}
{T}he classical problem of foam film rupture dynamics has been investigated when surfaces exhibit very high rigidity due to the presence of specific surfactants. 
Two new features are reported. First a strong deviation to the well-known Taylor-Culick law is observed. Then, crack-like patterns can be visualized in the film; these patterns are shown to appear at a well defined deformation. The key role of surface active material on these features is quantitatively investigated, pointing the importance of surface elasticity to describe these fast dynamical processes, and thus providing an alternative tool to characterize surface elasticity in conditions extremely far from equilibrium. The origin of the cracks and their consequences on film rupturing dynamics are also discussed.

\end{abstract}

\section{Introduction}
Despite its apparent useless character and simplicity, the dynamics of bursting of soap bubbles have fascinated scientists for more than a century. Lucien Bull (1904)  made the first images of soap bubble bursts. The first theoretical analysis dates back to Dupr\'e and then to \citet{TAYLOR1959} and \citet{CULICK1960} where they considered the presence of a rim at the edge of a hole created in the liquid film, collecting the liquid during its movement. The { constant} hole opening velocity $V_c$ results from a balance between {the rim} inertia and surface tension in the {film}, and is given by $V_c=\sqrt{2\gamma_\text{eq}/(\rho h_0)}$, with $\gamma_\text{eq}$ the {equilibrium} surface tension, $\rho$ the {liquid} density and $h_0$ the film thickness. These results are in good agreement with stationary experiments performed on liquid sheet \citep{TAYLOR1959} and has been extensively investigated by \citet{MCENTEE1969} in the case of soap films  {thicker} than 50 nm.
More recently, satellite formation during edge retraction \citep{Lhuissier2009} and bubble entrapment \citep{Bird2010} have been investigated as these behaviors are crucial in many applications. Destabilization of liquid sheets or bubbles indeed arise in many practical situations ranging from the building material industry, when glass sheets are molded,  to foam engineering, food processing, bio{logical} membrane and environmental science \citep{Bird2010}. 
In these applications, liquids can be viscous or contain surface active materials. 
{In the latter, surface tension becomes a dynamic{al} quantity{,} which depends on the local surface concentration of surfactants, and thus on the elongation of the surface; this {is} characterized by the surface elasticity defined as the derivative of surface tension with respect to relative changes in surface area.}
The effect of surface elasticity has been observed through the development of an aureole {surrounding} the opening hole {and expanding with time} \citep{FLORENCE1972,Liang1996,Lhuissier2009a}.
However, except in the case of {very} viscous liquid, the opening dynamics {always obey} Taylor-Culick law, {although}
some deviations have been reported {by} Mysels {\citep{MCENTEE1969, Florence1974}}, but hardly commented. 
In this work, {we investigate the dynamics of bursting of circular foam films {generated from} surfactant solutions inducing large surface elasticities and} we report for the first time {systematic} deviation{s} to Taylor-Culick law. A careful analysis allows us to {estimate surface elasticity at {both} large compression and compression rate in good} agreement with reported data in the literature.
Moreover, unexpected effects of frame size are observed {through} the appearance of new patterns, {reminiscent} of fractures or wrinkles in the film.

\begin{figure}
\centering	
\centerline{\includegraphics[width=.65\columnwidth]{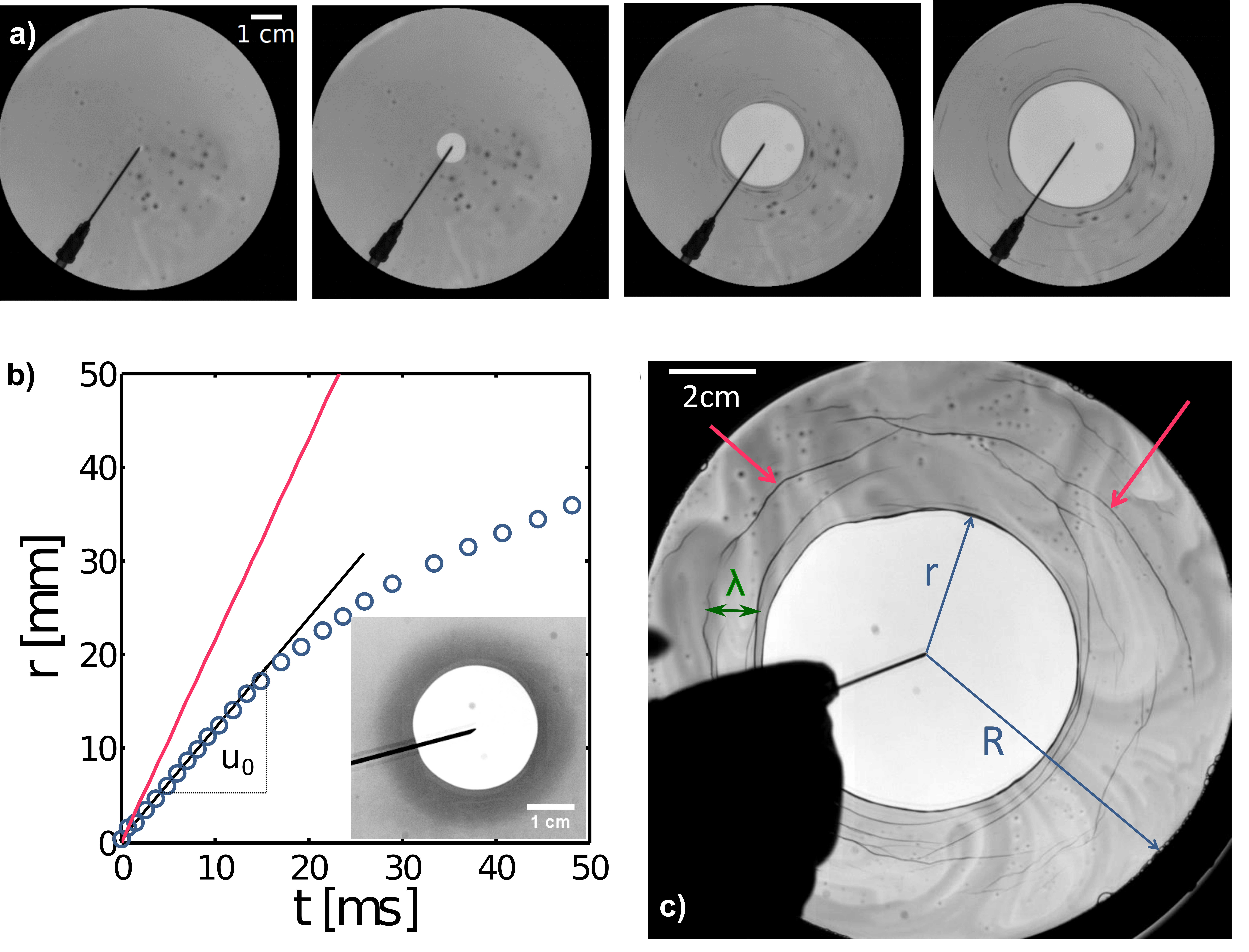}}
\caption{(a) Image sequence of a foam film rupture ($h_0$=10~$\mu$m, solution E -- table \ref{tab:2}). The timelapse between images is 8 ms. 
(b) Radius $r$ of the hole \emph{vs} time $t$ ($h_0$=10~$\mu$m, solution E). The {red (light)} line represents the prediction of Taylor-Culick, {while the black line shows the initial opening at constant velocity $u_0$}. Inset: Picture of a ruptured foam film (solution C) which highlights aureole formation.
(c) Picture of a foam film (solution E) 37 ms after its breaking. The arrows highlight {crack-like} {patterns}{,} which appear during the hole opening. 
}
\label{fig1}
\end{figure}


\section{Experimental set-up}

The experimental set-up {consists in a circular metallic frame of radius $R=1.5-11$~cm} pulled out from a surfactant solution at different velocities to generate films with various thicknesses. The film absolute thickness is determined through an absorption technique measurement \citep{Lastakowski2014, Petit2015} and we denote $h_0$ the initial average thickness of the film.  {Film rupture is initiated} by approaching a heated needle and {is} recorded via a high-speed camera (10000~Hz, Photron SA-4). An image sequence is reported in figure~\ref{fig1}a, where we  measure the radius $r$ of the expanding hole versus time, as shown in figure~\ref{fig1}b. Surfactant solutions are produced in  {a 10\%-90\% glycerol-water mixture} {in which}  a dye ({Brilliant Black BN  60\%, Sigma, 5g/L}) is added. They contain 3.3 g.L$^{-1}$ of sodium lauryl-dioxyethylene sulfate (SLES, {Stepan}), 1.7 g.L$^{-1}$ of cocoamidopropyl betaine (CAPB, {Goldschmidt}) and myristic acid (MAc, {Fluka}) in the concentrations $C$ described in table \ref{tab:2}. The surface elasticities  of similar solutions 
are well characterized in the literature \citep{Mitrinova2013} and span over two orders of magnitude when {the concentration $C$ of MAc is varied as reported in table \ref{tab:2}. {Such elastic moduli are attributed to the surface properties of the adsorbed layer of MAc, whose surface concentration is expected to increase with $C$ up to the saturation of the surface \citep{Golemanov2008}. At the same time, micelles of the two co{-}surfactants (SLES and CAPB) help to solubilize the poorly soluble fatty acid.}

	\begin{table}
  \begin{center}
\def~{\hphantom{0}}
  \begin{tabular}{lccccccc}
   Solution  & & A   	&   B 	& C	& D & E	& F	\\[8pt]
$C$ [mM] & & -				&0.055				&0.11		&0.22			&0.88		&2.2\\
$\gamma_\text{eq}$ [mN/m] &\hspace{15pt} & 29		&29			&27		&26			&23		&22\\[8pt]
$E_\text{od}$ [mN/m] & \hspace{15pt} & $4$	&$50$			&$90$	& $200$				&$400$	&$400$\\
$E_0 (u_0)$ [mN/m] & \hspace{15pt} &{60}& {200} & {2000} & {5000}	 & {$2.10^4 $}& {$4.10^4 $}\\
$E_0 (\text{cracks})$  [mN/m]   &\hspace{15pt} &  -  &  -  &  $90$  &  $200$  &  $300$  &  $300$  \\
  \end{tabular}
 \caption{Properties of the surfactant solutions used in the experiments: MAc concentration $C$, equilibrium surface tension $\gamma_\text{eq}$ and surface elasticities. Data from \citet{Mitrinova2013} for similar solutions (without glycerol and dye) are reported for $\gamma_\text{eq}$ and the elastic modulus $E_\text{od}$ measured with the oscillating drop method for small deformation ($0.2-4$ \%) at frequency 0.2 Hz. $E_0 (u_0)$ corresponds to the elasticity deduced from the initial hole  velocity using equation (\ref{eq}). $E_0 (\text{cracks})$ corresponds to the elasticity deduced from  cracking radius  using equation (\ref{eq2}).}
  \label{tab:2}
  \end{center}
\end{table}

\section{Results}
Some remarkable features can be underlined. At first, the opening velocity is constant  {as expected} but {smaller than predicted by} Taylor-Culick law (figure~\ref{fig1}b). Moreover, an aureole already described in the past \citep{FLORENCE1972,Liang1996,Lhuissier2009a} is observed {through spatial variations of transmitted light}, especially for the less rigid interfaces (inset of figure  \ref{fig1}b)}. Then, some dark patterns  are observed (see arrows in figure~\ref{fig1}c), which we denote \emph{cracks} in the following. This apparition coincides with a decrease of the opening velocity (figure~\ref{fig1}b), the presence of these cracks modifying the bursting dynamics.

\begin{figure}
\centering	
\centerline{\includegraphics[width=.8\columnwidth]{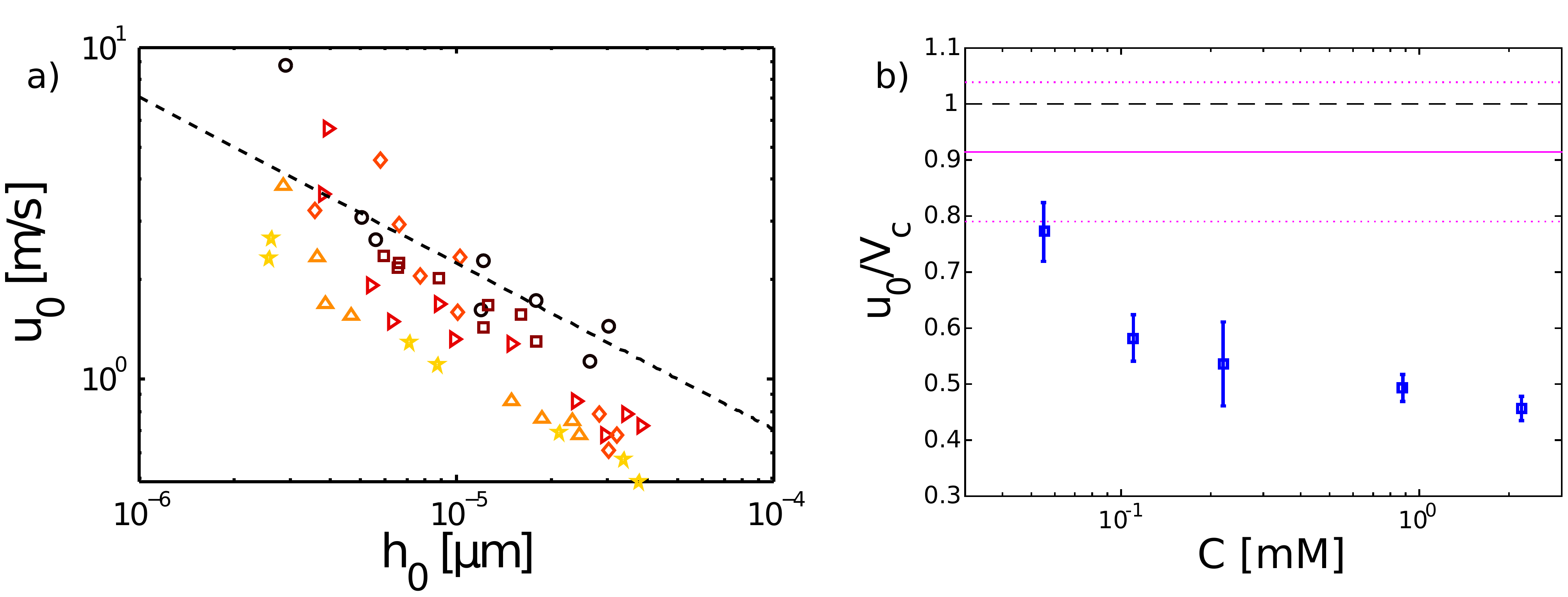}}
\caption{(a) Initial opening velocity $u_0$ of the {hole} as a function of the film thickness $h_0$ {for $R = 3$~cm}. The MAc concentration $C$ decreases from dark to light points: solutions A (o), B ($\square$), C ($\triangleright$), D ($\diamond$), E($\bigtriangleup$) and F ($\star$). (b) initial opening velocity normalized by Culick velocity $ u_0/V_c$ as a function of $C$ {(error bars: 95~\% confidence intervals). The magenta solid line shows the value measured for $C=0$, with error bars shown by the dotted lines. The values $u_0/V_c$ are extracted by performing least square percentage fit for each solution, with weights taking into account the 1~$\mu$m error on thickness measurements}.  In both figures, the black dashed lines represent Taylor-Culick law.}
\label{fig-VitIni}
\end{figure}

The initial opening velocity $u_0$ is represented in figure~\ref{fig-VitIni}a as a function of the initial film thickness $h_0$  {for various solutions}. Without MAc (solution A of table \ref{tab:2}), the velocity follows Taylor-Culick law ($\circ$), which is consistent with interfaces of low elasticity. However, in the presence of MAc, the initial velocity is lower than in the previous case. For each MAc concentration, the initial velocity varies with $1/\sqrt{h_0}\propto V_c$. For each solution {and different thicknesses}, we thus {extract} the initial opening velocity normalized by  Culick velocity. This quantity decreases when the MAc concentration increases (figure~\ref{fig-VitIni}b), {that is}, for larger surface elastic moduli \citep{Mitrinova2013}. 

\begin{figure}
\centering	
\centerline{\includegraphics[width=.8\columnwidth]{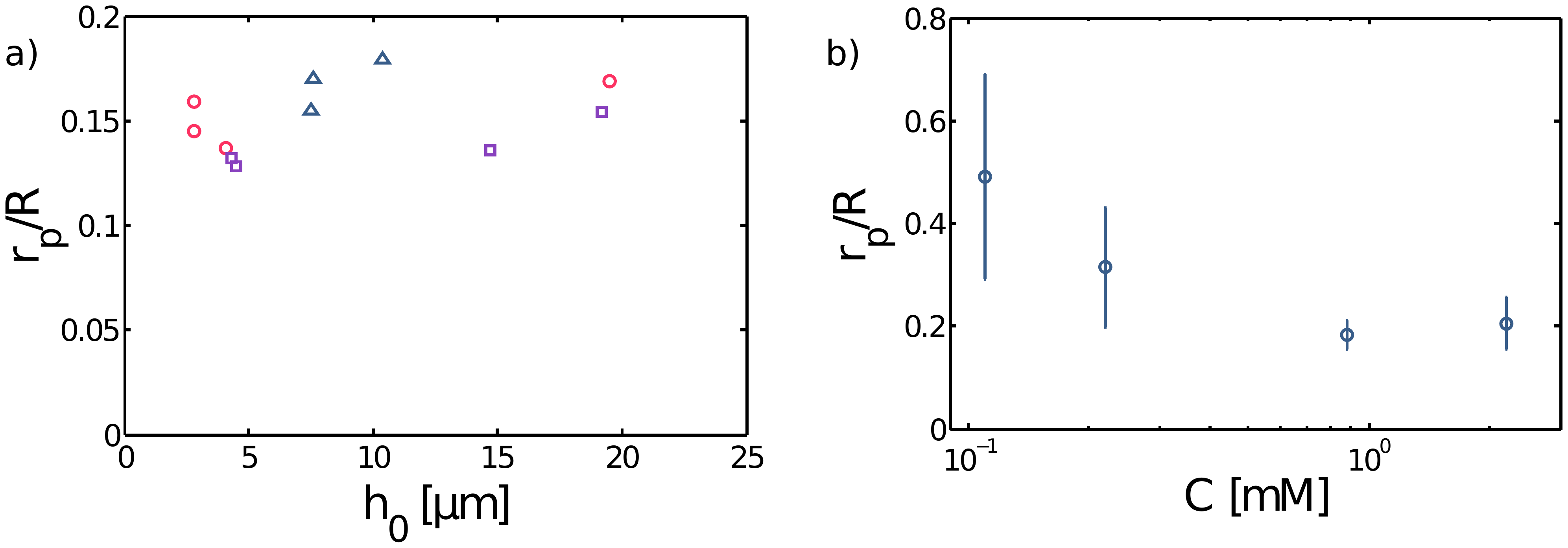}}
\caption{(a) Critical compression of the interface for crack formation $r_P/R$ as a function of the initial thickness $h_0$ for solution E and {different frame radii $R =1.5$~cm (o), $R =3$~cm ($\square$) et $R =11$~cm ($\bigtriangleup$). (b) $r_P/R$ averaged for thicknesses $h_0 = {2-35}$~$\mu$m and $R = 3$~cm, as a function of MAc concentration $C$.}}
\label{fig-RComp}
\end{figure}

During the film opening, {orthoradial} cracks {(perpendicular to the direction of opening)} appear in the film ({figures~\ref{fig1}a and \ref{fig1}c}), at a well defined radius of the hole $r_p$. {Some {specific} irregular fold-like patterns {and filaments} {have been  previously  reported} by \citet{MCENTEE1969}, although {not directly comparable to our observations.}} {For a given solution, figure~\ref{fig-RComp}a shows that the ratio $r_p/R$ is  independent of the frame radius (for $R=1.5-11$~cm) and almost independent of the film thickness (for $h_0 = 2-20$~$\mu$m). The cracks thus appear for a well-defined critical compression of the interface}. Figure~\ref{fig-RComp}b shows that this critical compression decreases with MAc concentration and the surface modulus. 

\section{Discussion}

These two observations concerning the initial opening velocity and the {onset compression for cracks} can be rationalized following the framework {initially} proposed by  \citet{FRANKEL1969} {for the {theoretical} description of aureoles}. 
{T}hey considered that surfactants are insoluble, which is reasonable at the timescale considered here: {the duration of the opening  $R/u_0$, typically 30~ms, is smaller than surfactant desorption time $\tau$. Indeed, although these processes are likely to be dominated by surfactant exchange with micelles in our systems \citep{Golemanov2008}, a lower bound for $\tau$ is provided by the diffusion time across the film thickness $h_0^2 / D \approx 40\text{~ms}-2\text{~s}$ (for $h_0 = 2-40$~$\mu$m and $D = 10^{-10}$~m$^2$/s).} {Adsorption times longer than 30~ms for myristic acid in {these} systems have also been reported \citep{Mitrinova2013b}}. {A compressive shock thus} propagates at the surface of the film. The liquid is  collected in an extended rim ---~an $aureole$~--- {visible in figure~\ref{fig1}b (inset) and} whose shape depends on surface tension, film thickness and surface elasticity. 

{Viscous effects have also been neglected. Indeed, as no shear takes place within the film thickness, the characteristic Reynolds number and \textit{surface} Reynolds numbers
read $\Rey= u_0 R / \nu  \gg 1$   and  $\Rey_s =\rho  u_0 R h_0 / \kappa $, respectively, with $\nu$ the kinematic bulk viscosity and $\kappa$  the intrinsic surface viscosity. Surface viscous dissipation can \textit{a priori} not be neglected if values of $\kappa$ measured at 0.2 Hz are considered {\citep{PhDCosta,Golemanov2008}}. However, surface viscosity is expected to collapse at large frequencies, as shown in experiments and modeling {\citep{Lucassen1972}.} Eventually, the observation of a constant initial velocity varying with $1/\sqrt{h_0}\propto V_c$ (figures~\ref{fig1}b and \ref{fig-VitIni}) is a key indication that inertia (and not viscous effects) is dominant in this problem.}
%
\subsection{Deviation to Taylor-Culick law}
The dynamics of the rim is then controlled by the balance between inertia and surface tension {spatial} gradient. 
We {assume here that} surface elasticity is constant up to a certain compression. 
{In this {particular} case}, the velocity of the aureole front (delimiting the {frontier with the zone of undisturbed film whose thickness is still $h=h_0$}) simply reads
$u_{f}= \sqrt{2 E_0/(\rho h_0)}= V_C\sqrt{{E_0}/{\gamma_\text{eq}}}$, {which can be seen as a two dimensional analogous of sound (compression) velocity}.
The opening hole velocity can also be determined by solving the self-similar profile of the aureole and {applying} mass conservation. No analytical solution is provided in the considered radial geometry but numerical resolution shows that 
\begin{equation}
u_0=V_c \ {f}(E_0/\gamma_\text{eq})
\label{eq}
\end{equation}
with ${f}$ a decreasing function determined numerically (see appendix~\ref{appA}) and reported in figure~\ref{figure4}. 
It is thus still proportional to Taylor-Culick velocity $V_c$ and decreases with the interfacial elasticity $E_0$, which {is consistent} with experimental observation{s} of {figure}~\ref{fig-VitIni}.
{From these data and equation (\ref{eq})}, an interfacial elasticity {$E_0(u_0)$} can be deduced ({figure}~\ref{figure4}), {which} is reported in table \ref{tab:2} as a function of the MAc concentration. These data are compared to {measurements of surface moduli $E_\text{od}$} from the oscillating drop method {performed} by  \citet{Mitrinova2013}. It shows the same qualitative variation {with $C$ despite} a discrepancy on the {absolute} values obtained. However, the shrinkage amplitude and the compression timescales differ by several orders of magnitude, and the surfactant monolayer {at the} interface is expected to be highly non-Newtonian \citep{PhDCosta, Lucassen1972}.

\begin{figure}
\centering	
\centerline{\includegraphics[width=.5\columnwidth]{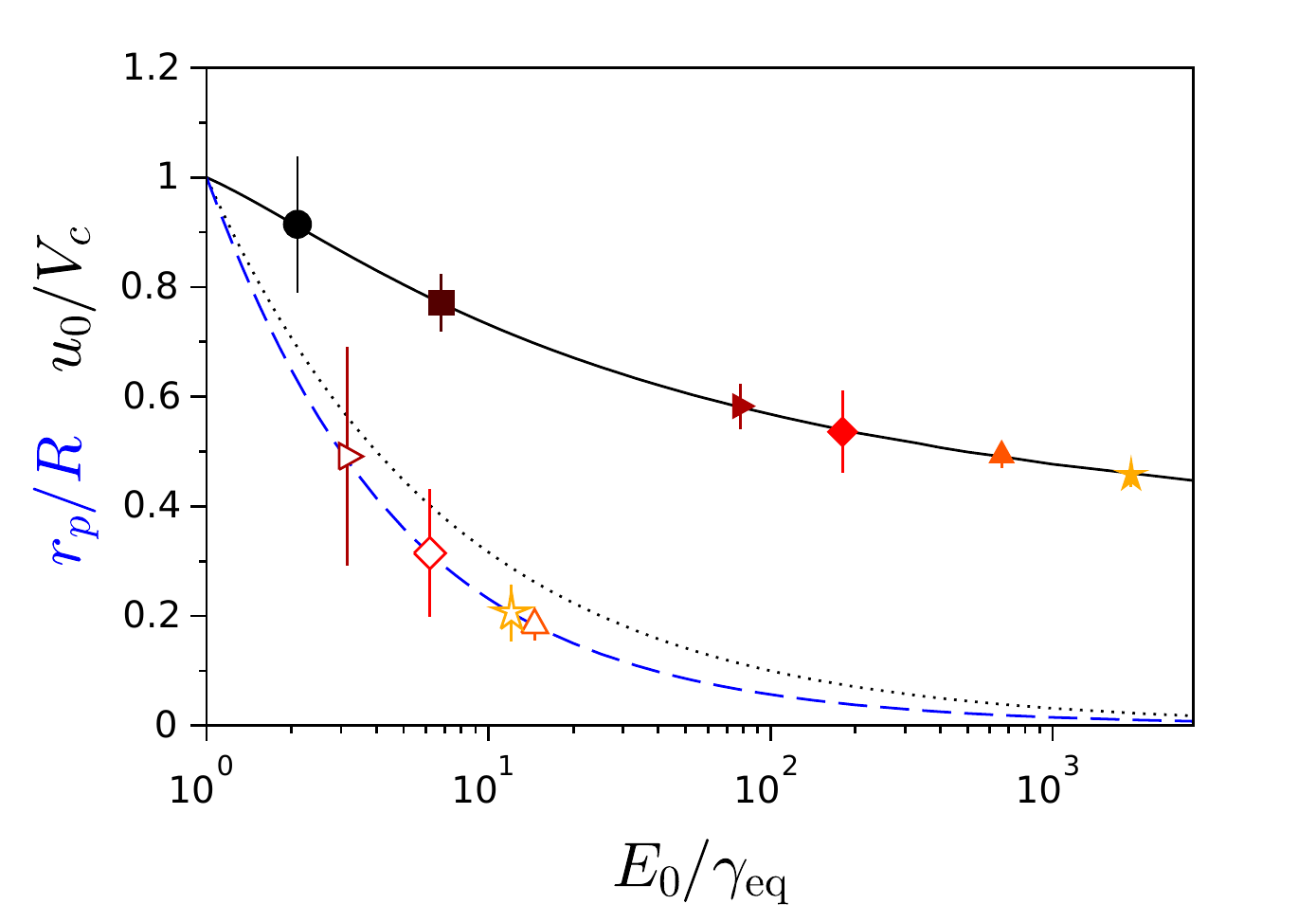}}
\caption{Solid black line: numerical prediction for  the normalized hole velocity $u_0/V_c = {f}(E_0/\gamma_\text{eq})$ for radial bursting (equation (\ref{eq}) and  \ref{appA}). {The dashed line corresponds to $u_0/V_c =\sqrt{\gamma_\text{eq}/E_0}$ expected {for unidimensional bursting} \citep{FRANKEL1969}.} Blue dashed line: prediction for the critical radius at which cracks appear  $r_P/R = \sqrt{{\gamma_\text{eq}}/{E_0}}f({E_0}/{\gamma_{eq}})$ (equation (\ref{eq2}). The solid (resp. empty) symbols correspond to the experimental data from figure~\ref{fig-VitIni} (resp. figure~\ref{fig-RComp}), from which we determine the elastic moduli $E_0(u_0)$ (resp. $E_0(\text{cracks}$)). {S}ame symbols and colors as in figure~\ref{fig-VitIni}a.}
\label{figure4}
\end{figure}

\subsection{Crack appearance}
{Besides, }snapshot inspection shows that cracks appear when the compressive surface wave (i.e. the aureole front) reaches the metallic frame of the film. 
{Cracks are thus expected for:}
\begin{equation}
\frac{r_p}{R}=\frac{u_0}{u_f}=\sqrt{\frac{\gamma_\text{eq}}{E_0}}f\left(\frac{E_0}{\gamma_\text{eq}}\right)
\label{eq2}
\end{equation}
This prediction, represented in {figure}~\ref{figure4}, is indeed in good agreement with our observations: The hole radius when cracks appear $r_p$ increases with the frame radius $R$ and decreases with surface elasticity probed through MAc concentration variations, as shown in figure~\ref{fig-RComp}. Eventually,  this critical compression does not depend on film thickness $h_0$, showing that elasticity is not affected by confinement  in the experimental configuration tested.

The surface elasticity {$E_0 (\text{cracks})$} can therefore be deduced from the critical radius for crack apparition $r_p$ {(figure~\ref{figure4})}, and reported for the different MAc concentration in table \ref{tab:2}. In this case, a very good agreement is obtained with {the measured value of} the surface modulus \citep{Mitrinova2013}, which confirms that the cracks {arise} from a compression of the aureole when its front reaches the frame. 

{Note that the values of surface elasticity deduced from our two methods may differ. This is however expected due to our strong hypothesis {of constant elasticity}. Indeed, {while the} aureole front velocity only depends on the surface elasticity at very low compression rate (at the edge of the undisturbed film), the hole opening velocity modeling takes into account the elasticity through large interface compression. {For large deformation}, it is expected that the constant elasticity model fails{: at large compression, the myristic acid surface concentration increases, which should result in larger elasticity as can be inferred from the moduli dependency upon $C$ \citep{Mitrinova2013}}. {The effective modulus $E_0(u_0)$} should then deviate more from measurements {at small deformations} performed by the oscillating bubble technique \citep{Mitrinova2013}.

{In addition, }the effect of elasticity has indirect consequences of some features of foam film rupture. For example, {no {flapping nor} transverse destabilization of the rim was observed for our rigid soap films, in contrast {to} observations on low elasticity films and theoretical predictions \citep{Lhuissier2009}; {however, the reduced rim velocity could prevent the flapping instability to develop and subsequent film atomization \citep{Lhuissier2009}}. 

\subsection{Crack-like patterns}
Let us now discuss  the {observed} crack{-like} patterns. During the fast deformation of the surface, the surfactants behave as an insoluble monolayer, {comparable} to {a} lipid monolayer experiencing a compression in a Langmuir trough \citep{Lee2008}. In this case, above a critical compression, such a monolayer can behave differently depending on its structure. If it is liquid-like, it ejects the molecules in the bulk in the form of vesicles or bilayers. If it is solid-like, it can bend as an elastic sheet or fracture as a fragile material. 

Although {our experiment does not provide a microscopic characterization of this transient surface structure}, 
{the crack pattern can be macroscopically characterized}. In particular, even though the cracks are irregularly {distributed}, a number of cracks per  radial segment can be counted;  the deduced characteristic length between two cracks  denoted $\lambda$ (figure~\ref{fig1}c)  is reported in figure~\ref{fig-AL} as a  function of MAc concentration $C$ (a) and film thickness $h_0$ (b). 
\begin{figure}
\centering	
\centerline{\includegraphics[width=.98\columnwidth]{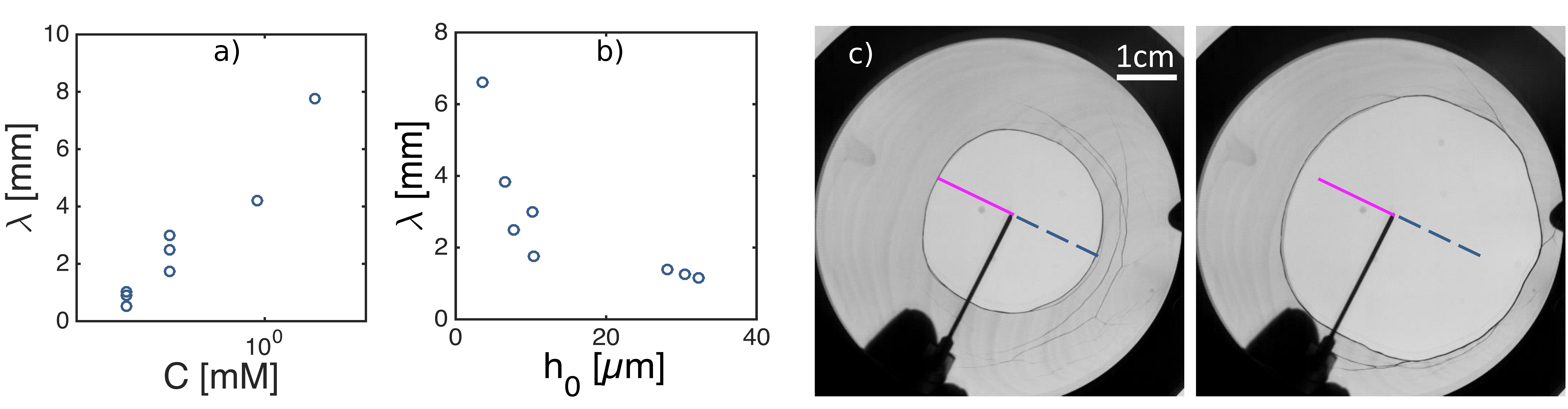}}
\caption{(a) Characteristic length $\lambda$ between two cracks as a function of MAc concentration $C$  for $h_0 =  11\pm3$~$\mu$m. (b) $\lambda$ as a function of the film thickness $h_0$ for solution D. (c) Bursting of a soap film of thickness $h_0 = 3$~$\mu$m (solution E). The timelapse between the two images is 9~ms, and the two lines highlight the velocity inhomogeneities.}
\label{fig-AL}
\end{figure}

The {increase of $\lambda$ with $C$} is expected whatever the mechanism proposed. On the one hand, for higher bulk concentration, {solubilization of interfacial surfactants is more difficult, hence a reduced number of vesicles or bilayers are {to} be expelled.} 
On the other hand, a more concentrated solid-like layer will also exhibit a higher bending modulus {and wavelength of elastic ripples are expected to increase with this modulus \citep{Cerda2003}}.
The {decrease of the characteristic length with the film thickness $h_0$} is more unexpected.  {{For} the solid-like behavior, a thinner elastic sheet will bend more easily than a thicker one, thus exhibiting smaller ripple wavelength when buckled \citep{landau1975elasticity}, in contrast with our observations.}
{If cracks correspond to monolayer collapse by vesicles formation, it should not be affected by the film thickness}. 
However, when modifying the thickness of the film, we also vary the velocity of compression or shrinkage rate. This parameter induces dynamical structural change in the surfactant monolayers (as it does in {bulk} crystallization processes for example \citep{Cabane2003}). 

{Finally,} a complete understanding of the origin of these crack-like patterns would require some local high speed imaging structural analysis, which are beyond the scope of the present paper.

The presence of these irregular cracks have direct consequences on hole opening dynamics. Indeed, when the aureole {reaches} the metallic frame, the hole opening slows down (and even stops for the thinner rigid films) and then irregularly accelerates in the region where the cracks appears. This feature is reported in figure~\ref{fig-AL}c. Moreover, a velocity discontinuity in the liquid is observed, the {outer} region being at rest whereas the {inner} region {is deformed.} 

\section{Conclusion}
{To conclude,} we have shown that modifying the chemistry of surfactant solutions can have strong influences on macroscopic dynamical processes, as observed in various situations {in foams and foam films} \citep{Couder1989, Durand2006, Seiwert2013, Petit2015, Lorenceau2009}. 
However, we have investigated here this effect {under large deformations and} in a {fast} dynamical process, \textit{i.e.} at large Reynolds numbers, where {the effects of molecular scales and surfactants} are expected to be negligible. 

{The initial constant velocity} opening dynamics is well described taking into account the surface elasticity of the interfaces {and was shown to {be reduced} at high surface modulus. This may be responsible for the inhibition of rim fragmentation and droplet ejection usually reported in liquid film ruptures \citep{Lhuissier2009}. Further studies should determine the role of the ejected droplets in rupture propagation in macroscopic foams; the stability of these systems is indeed known to depend dramatically on the surface elastic properties \citep{Rio2014a}.}
However, finite size effects becomes soon crucial: when the elastic compression surface wave reaches the border of the frame, crack-like patterns, where velocity discontinuity are observed, appear in the foam film. Determining the origin of cracks, their microscopic structure, their location and number, and how they control film opening dynamics remain a challenge to tackle.\\

The authors thank Gilles Simon for his help in setting up the experiment.

\appendix
\section{}
\label{appA}

\subsection*{Equations for radial bursting}

We describe the radial bursting dynamics of a foam film of initial uniform thickness $h_0$ and include the effect of dynamic surface tension as first proposed by  \citet{FRANKEL1969}: the surface tension $\gamma$ is assumed to depend only on the shrinkage of the surface $\alpha$ which by mass conservation is related to film thickness $\alpha=  h_0/h$. We denote the surface elasticity $E(\alpha) = \frac{\mathrm{d}\gamma}{\mathrm{d} \alpha}$. 
As viscous terms are negligible, the capillary forces are balanced by the fluid inertia. Variations of fluid velocity  {across the film} are also neglected and equations are averaged over $h$. {These equations can be explicitly solved in the unidimensional case \citep{FRANKEL1969}. However, in the case of radial bursting, a numerical resolution is necessary.}

We consider a material element that has initially the position $R$ (\emph{i.e.} that has Lagrangian variables $(R,t)$) . At instant $t$, its position is $r(R,t)$ {and its thickness $h(R,t)$} (figure~\ref{fig:appendix}a). 
The fluid velocity is $u  = \partial{r}/\partial{t}$ and the shrinkage is defined as $\alpha = {h/h_0} \partial{r^2}/\partial{R^2} = (r/R) \partial{r}/\partial{R}$. The momentum balance on the fluid element yields
$$
\rho r h \frac{\partial{u}}{\partial{t}} = 2 r \frac{\partial{\gamma}}{\partial{r}} 
$$
which can be rewritten
\begin{equation}
\label{eq1}
\frac{\partial{u}}{\partial{t}}  =   \frac{2 E(\alpha)}{\rho h_0} \frac{r}{R} \frac{\partial{\alpha}}{\partial{R}} = U_\alpha^2 \frac{r}{R} \frac{\partial{\alpha}}{\partial{R}} 
\end{equation}
in which we have defined the characteristic velocity
$$U_{\alpha} = \sqrt{\frac{2 E(\alpha)}{\rho h_0}}.$$

Following the analysis of \citet{FRANKEL1969}, we are looking for self-similar solutions in the form $r/t  = f(R/t)$. We define the variables $W = R^2/(2t^2)$ and $w = r^2/(2t^2)$ ($w$ and $W$ have the dimensions of square velocities) and we expect $w = w (W)$. The relative shrinkage is also set by $\alpha= \mathrm{d}w/\mathrm{d}W$. Starting from equation (\ref{eq1}), we find
\begin{equation}
 {\frac{W}{ w}}  \left [ 1  -  \frac{W}{w} \frac{\mathrm{d}w}{\mathrm{d}W}    \right ] \frac{\mathrm{d}w}{\mathrm{d}W}   = \left [ U_{\alpha = \mathrm{d}w/\mathrm{d}W}^2  -  {\frac{2W^2}{ w}} \right ] \frac{\mathrm{d}^2 w}{\mathrm{d} W^2}
\label{eq2bis}
\end{equation}


\begin{figure}
    \centering
    \centerline{\includegraphics[width=.98\columnwidth]{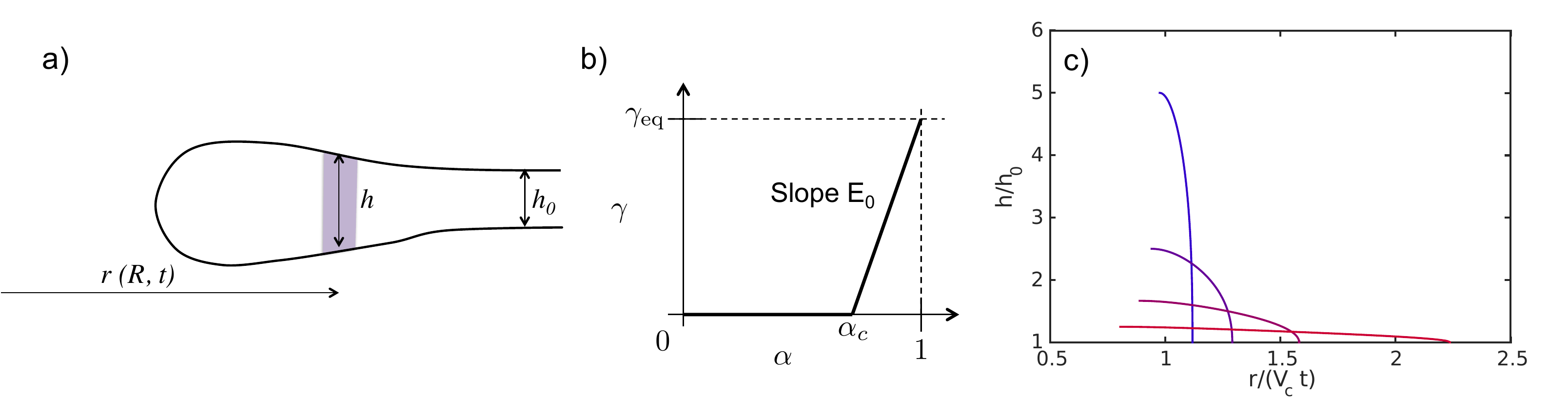}}
\caption{(a) Profile of the film and notations. (b) Variations of surface tension $\gamma$ versus shrinkage $\alpha$ in the simplified constant elasticity modeling. (c) $h/h_0$ as a function of $r/(V_c t) = \sqrt{2 w/V_c^2}$ for radial bursting and different values of $\alpha_C$ (0.2, 0.4, 0.6, 0.8 from top to bottom at the origin).}
\label{fig:appendix}
\end{figure}

 {A first information on the film dynamics can be inferred from this equation: Far from the hole, i.e. for large $W$, the film should remain undisturbed, which corresponds to $w = W$ and $\mathrm{d}w/\mathrm{d}W = 1$. This condition combined with equation (\ref{eq2bis}) yields $\left [ U_{\alpha=1}^2  -  {\frac{2W^2}{ w}} \right ] \frac{\mathrm{d}^2 w}{\mathrm{d} W^2} = 0$, which implies that the matching with the disturbed film can only be done at $W = W_0 = U_{\alpha=1}^2/2$. The velocity of the front of the \emph{aureole}, or extended rim corresponding to the disturbed film, is thus given by $u_f = U_{\alpha=1}$ \citep{FRANKEL1969}.}

{Finally, the complete aureole profile and hole receding velocity} will depend on the form of the elasticity versus shrinkage. 

\subsection*{Numerical resolution for a constant elasticity model}

We consider at first order a model of constant elasticity $E_0$, as described in figure~\ref{fig:appendix}b. We introduce here $\alpha_c$, which corresponds to the maximum shrinkage the film can endorse. For $\alpha > \alpha_c$, the surface elasticity is constant and reads
$$
\frac{\mathrm{d} \gamma}{\mathrm{d}\alpha}  = E_0 =\gamma_\text{eq} \frac{1}{1-\alpha_c}
$$ and then 
$$
U_{\alpha} = U_0 = \sqrt{\frac{2 E_0}{\rho h_0}} = \sqrt{\frac{E_0}{\gamma_\text{eq}}} \ V_c
$$
where $V_c = \sqrt{2 \gamma_\text{eq}/(\rho h_0)}$ is Culick velocity.


When $\alpha>\alpha_c$, equation (\ref{eq2bis}) can be written as
\begin{equation}
 {\frac{W}{ w}}  \left [ 1  -  \frac{W}{w} \frac{\mathrm{d}w}{\mathrm{d}W}    \right ] \frac{\mathrm{d}w}{\mathrm{d}W}   = \left [ U_0^2  -  {\frac{2W^2}{w}} \right ] \frac{\mathrm{d}^2 w}{\mathrm{d} W^2}
\label{eq:radial_dimension}
\end{equation}
In non-dimensionalized form (stating $\tilde{W}=2W/V_c^2$ and $\tilde{w}=2w/V_c^2$), this equation reduces to
\begin{equation}
 {\frac{\tilde{W}}{\tilde{ w}}}  \left [ 1  -  \frac{\tilde{W}}{\tilde{w}} \frac{\mathrm{d}\tilde{w}}{\mathrm{d}\tilde{W}}    \right ] \frac{\mathrm{d}\tilde{w}}{\mathrm{d}\tilde{W}}   = \left [\frac{1}{1-\alpha_c}  -  {\frac{2 \tilde{W}^2}{\tilde{w}}} \right ] \frac{\mathrm{d}^2 \tilde{w}}{\mathrm{d} \tilde{W}^2}
\label{eq:radial_nodimension}
\end{equation}
with the two following boundary conditions. In $\tilde{W}=0$, at the hole, we have the maximum shrinkage (minimum value of $\alpha_c$): $\frac{\mathrm{d}\tilde{w}}{\mathrm{d}\tilde{W}} (\tilde{W}= 0) = \alpha_c$. In $\tilde{W}= \tilde{W}_0 = \frac{1}{2(1-\alpha_c)}$, at the aureole front, the solution should match the undisturbed film solution $\tilde{w} (\tilde{W}_0) = \tilde{W}_0 $.

This equation is solved numerically 
 with a shooting method. 
From the function $w(W)$, we can deduce the thickness profile, using the relation $h/h_0 = 1/(\mathrm{d}w/\mathrm{d}W)$   for different elasticities (figure~\ref{fig:appendix}c). As the elasticity increases (i.e. as $\alpha_c$ becomes closer to 1), we find that the aureole is thinner and wider{, while for $E_0 = \gamma_\text{eq}$ (corresponding to $\alpha_c = 0$), one recovers a punctual rim receding at Taylor-Culick velocity $V_c$}.

We can also estimate the initial hole velocity $ u_0 = \sqrt{2 w(W=0)}$, which is shown in figure~\ref{figure4} as a function of the ratio $E_0/\gamma_\text{eq}$. 
We also observe that the results obtained deviate from {those obtained for unidimensional bursting \citep{FRANKEL1969}},  especially for large elasticities, {emphasizing}  the crucial role of radial geometry.

\bibliography{litThese}
\bibliographystyle{jfm}

\end{document}